\documentclass[11pt]{article}
\usepackage{graphicx}
\newcommand{\be}{\begin{equation}}
\newcommand{\ee}{\end{equation}}

\newcommand{\bea}{\begin{eqnarray}}
\newcommand{\eea}{\end{eqnarray}}

\newcommand{\mbold}[1]{\mbox{\boldmath$#1$}}
\newcommand{\bnabla}{\mbold{\nabla}}
\begin{document}
\title{\bf
QUASI-LOCAL DENSITY FUNCTIONAL THEORY\\
AND ITS APPLICATION WITHIN EXTENDED\\
THOMAS-FERMI APPROXIMATION}
\author{V. B. Soubbotin and V. I. Tselyaev \\
{\it \small Nuclear Physics Department,
 V. A. Fock Institute of Physics},\\
{\it \small St. Petersburg State University, 198504,
 St. Petersburg, Russia}\\$\vphantom{,}$\\
 X. Vi\~nas \\
{\it \small Departament d'Estructura i Constituents de
 la Mat\`eria, Facultat de F\'{\i}sica},\\
{\it \small Universitat de Barcelona,
 Diagonal 645 E-08028, Barcelona, Spain}}
\date{}
\maketitle
\begin{abstract}
A generalization of the Density Functional Theory is proposed. The
theory developed leads to single-particle equations of motion
with a quasi-local mean-field operator, which contains a
quasi-particle position-dependent effective mass and a spin-orbit
potential. The energy density functional is constructed using the
Extended Thomas-Fermi approximation. Within the framework of this
approach the ground-state properties of the doubly magic nuclei
are considered. The calculations have been performed using the
finite-range Gogny D1S force. The results are compared with the
exact Hartree-Fock calculations.
\end{abstract}

\vspace{2em}
\newpage
\section{INTRODUCTION}

The Hartree-Fock (HF) method is one of the most widely used
approaches in nuclear physics. It is based on the concept of
independent particle motion in the mean field produced by
effective nucleon-nucleon forces which are generally non-local and
density dependent. The resulting equations of motion contain the
non-local single-particle potential (SPP) which is determined
self-consistently. The comprehensive study of nuclear ground state
properties within the HF method has been performed with the
zero-range Skyrme-like forces \cite{Vauth,Chab}. The status of
this problem is not completely the same for finite range forces.
The exact solution of the HF equations in this case is not an easy
task, mainly due to the non-locality of the SPP. For example, the
complete solution of the HF equations was carried out in
Ref.~\cite{Gog} for a finite range effective force with a Gaussian
formfactor using a harmonic oscillator basis. On the other hand,
the M3Y effective force with a Yukawa formfactor was employed in
Ref.~\cite{Hof} within the Campi and Bouyssy \cite{CB} local
approximation for the single-particle density matrix.

Thus the problem of localization of the non-local SPP becomes
actual. As it is known the non-local exchange Fock part of the SPP
is determined using the non-local single-particle density matrix
(DM in the following). If one approaches the DM in terms of only
local quantities such as the particle density and kinetic-energy
density, the corresponding HF exchange energy becomes a functional
of these local quantities. The equations of motion obtained from
the resulting local HF functional are second-order differential
equations. It is important to note that they do not contain any integral
operators which lead to difficulties in the general non-local
case. For instance the Negele and Vautherin expansion of the DM
\cite{NV} and its modification by Campi and Bouyssy \cite{CB}
enables the HF energy to be expressed in the pointed functional
form. Recently, another approach based on the Extended
Thomas-Fermi method (ETF, see, for instance, Ref.~\cite{RS}) has
been proposed to this aim in Ref.~\cite{SV}.

An alternative, in a sense, approach
to the mean field theory, which is
widely used in applications to electron systems, is based on
the Kohn-Sham (KS) \cite{KS} method within the framework of the
Density Functional Theory (DFT). The original version of this
theory was developed in the pioneering paper of Hohenberg and Kohn
(HK) \cite{HK} where an energy functional depending only on the local
particle density was considered (so we shall call it local
DFT). Later on other versions of the DFT were proposed
(see, for example, Refs.~\cite{JG,DFT}).
In particular, the non-local extension of this theory was discussed
by Gilbert in Ref.~\cite{Gilb} where the functional dependence on
the DM was included.

The main merit of the KS scheme consists of the following: it
gives a way to obtain single-particle equations of motion
for the local DFT. These equations contain the local mean-field
potential that has to be determined self-consistently. Notice that
in contrast to the approximate HF method, the DFT yields, in
principle, the \textit{exact} ground-state energies (and the
referred quantities) of the many-body system. Concerning the
single-particle spectrum only the last occupied level has the
exact physical meaning of the chemical potential in the DFT which
is just the particle separation energy.

There is one more important difference between the HF and KS
methods that is revealed in the applications to nuclei. The
radial-dependent effective mass and the spin-orbit potential are
essential components of the HF approach in nuclear physics.
These two quantities arise owing to the
kinetic-energy density and spin density dependence of the HF
energy functional. However, in the original KS method the
effective mass is constant and equal to the physical mass and
there is no spin-orbit potential because this method
starts from a local energy density functional.
At the same time at least the spin-orbit potential is
necessary for the realistic description of nuclear properties.
It is possible to introduce the kinetic-energy
density and spin density dependence of the DFT energy functional
formally. In this case, to derive the single-particle equations
following the ideology of the KS method, one would assume that any
kinetic-energy density and spin density can be produced by the
many-particle wave function describing the non-interacting system
in some external potential with a spin-orbit component. But,
as opposed to the case of the local particle density (see
Ref.~\cite{Lieb}), this statement \textit{has not been proved}.

To include the radial-dependent effective mass and the spin-orbit
potential in the consideration in a rigorous way, one would have to
use the non-local extension of the DFT and to derive
single-particle equations of motion directly from the energy
functional of this theory. However the straightforward application
of the standard variational principle to the non-local energy
functional leads to serious difficulties in view of the specific
properties of the pseudo-Hamiltonian obtained (see
Ref.~\cite{Gilb} for details).

The main goal of the present paper is to develop the modification
of the non-local generalization of the DFT which would be free
from the above-mentioned shortcomings of the non-local theory.
To this aim we define an energy
functional that depends on the DM produced by a determinant wave
function (in what follows we call it by a Slater-determinant DM).
Although this DM generally does not correspond to any interacting
fermion system nevertheless, we will show that the minimum of
this functional coincides with the \textit{exact} ground-state
energy of the interacting system under consideration.
Applying the variational principle, we derive the single-particle
equations of motion which, in contrast to the KS equations, contain
a non-local SPP. This is described in the second part.

In the third part the quasi-local reduction of the DFT is developed.
Within the quasi-local DFT the energy functional depends on the
local particle densities as well as on the \textit{uncorrelated}
kinetic-energy and spin densities. The single-particle equations,
which are obtained by the minimization of this functional, contain
the local SPP, the \textit{uncorrelated} radial-dependent
effective mass and the spin-orbit potential.

In the fourth part we derive a semiclassical HF energy functional
within the quasi-local scheme starting from the recently proposed
expansion of the DM in the Extended Thomas Fermi method \cite{SV}.
The explicit formulae for the energy functional, the SPP and the
effective mass are obtained using the finite-range Gogny force
\cite{Gog}, as described in the fifth part. The residual
correlation term is taken phenomenologically. In the sixth section
we apply our method to the description of the ground-state
properties of some doubly-magic spherical nuclei. The main results
are set out in the summary. In the Appendix 1 some auxiliary formulae for
the SPP are given. In Appendix 2 we describe a simple method
to take into account the two-body correction of the centre-of-mass
motion to the binding energy.

\section{THE NON-LOCAL GENERALIZATION OF THE DFT}

Let us consider a system of $N$ interacting fermions. In the
nuclear case we are interested in systems with two kinds of
particles, namely neutrons and protons. Let $H$ be the
non-relativistic many-particle Hamiltonian. The explicit form of
this operator is not important here. One can associate it with the
usual formulae:
\begin{equation}
H=T+\sum_{i\ne j}v^{NN}_{ij}+\sum_{i\ne j}v^{Coul}_{ij}
   + \cdots \,,
\label{H}
\end{equation}
where
\begin{equation}
 T = - \sum_i \frac{\hbar^2}{2m} \Delta_i
\label{ke}
\end{equation}
is the kinetic-energy operator, $v^{NN}_{ij}$ is the bare
nucleon-nucleon (NN) strong two-particle interaction, $v^{Coul}_{ij}$
is the Coulomb force acting between protons and the dots note the
many-particle interactions if needed.

The HK energy functional \cite{HK}, which only depends on the local
particle density $n$, can be defined within the framework of the
constrained search method as follows (see, for example, \cite{JG,DFT})
\begin{equation}
E_{HK}[n]=\inf_{\Psi\rightarrow n}<\Psi|H|\Psi>\,, \label{eq1}
\end{equation}
where $|\Psi>$ is an arbitrary normalized $N$-particle state. The
short notation $\Psi \rightarrow n$ hereinafter means the
many-to-one mapping of the wave function $\Psi(x_1,\ldots , x_N)$
to the local density $n({\mbold r})$ i.e. it means
that the following equalities are fulfilled
\begin{eqnarray}
n({\mbold r})&=&n_p({\mbold r})+n_n({\mbold r})\,,
 \label{eq2s}\\
 n_q({\mbold r}) &=& \sum_{\sigma} \rho (x, x)\,,
 \label{eq2}\\
 \rho (x, x') &=&
 N \int \Psi (x, x_2,\ldots , x_N) \Psi^* (x', x_2,\ldots , x_N)
 dx_2 \cdots dx_N\,,
 \label{defrho}
\end{eqnarray}
where $\rho (x, x')$ is the single-particle DM, $x=\{{\mbold
r},\sigma,q\}$ includes the spatial $\mbold{r}$ and spin $\sigma$
variables and the index of nucleon type $q = n, p$.
The integration over $x$ includes the
summation over $\sigma$ and $q$.

The functional (\ref{eq1}) depends on the total local density
$n({\mbold r})$. One can define other energy functionals which
are  dependent either on $n_q({\mbold r})$ or even on $\rho (x, x)$.
The particular choice of functional dependence is determined by
the task under consideration.

In the local DFT it is proved that the minimum of the functional
$E_{HK}[n]$ is just the \textit{true} ground-state energy $E_{GS}$
and is attained for the \textit{true} ground-state density
$n_{GS}$. To obtain $E_{GS}$ and $n_{GS}$, one can use the KS method
which yields single-particle equations similar to the HF equations.
The rigorous derivation of these equations is based on the following
statement proved by Lieb \cite{Lieb}:

\textit{If $n(\mbold{r}) \ge 0$, $\int n(\mbold{r}) d \mbold{r} =
N$, $\int (\bnabla \sqrt {n(\mbold{r})})^2 d \mbold{r} < \infty$,
then there exists $N$-particle Slater-determinant wave function
$\Psi_0$  built up from an orthonormal set of N single-particle
wave functions $\varphi_i$:}
\begin{equation}
 \Psi_0 (x_1, \ldots , x_N) = (N!)^{-1/2} 
 \mbox{det} \{ \varphi_i (x_j) \} \,,
\label{dets}
\end{equation}
\textit{such that $\Psi_0 \rightarrow n({\mbold r})$}. In other
words there is a many-to-one mapping of $N$-particle Slater
determinant wave functions onto the local particle density $n({\mbold
r})$.

 This theorem enables one to define the kinetic-energy functional
$T_0[n]$ for a system of non-interacting particles:
\begin{equation}
T_0[n]=\inf_{\Psi_0\rightarrow n}<\Psi_0|T|\Psi_0>\,, \label{eq3}
\end{equation}
and to divide the HK functional $E_{HK}[n]$ into two parts:
\begin{equation}
E_{HK}[n]=T_0[n]+W[n], \label{eq4}
\end{equation}
where the energy functional $W[n]$ contains the potential energy
as well as the correlation part of the kinetic energy.

Because the density $n$ is produced by some Slater-determinant
wave function one has from Eqs.~(\ref{eq2})-(\ref{dets}): \be
n_q({\mbold r})=\sum_{i=1}^N \sum_{\sigma} |\varphi_i({\mbold r},
\sigma, q)|^2\,. \label{eq4a} \ee
By the same reasoning the
kinetic-energy functional of the non-interacting system
(\ref{eq3}) can be written as
\be T_0[n] = \sum_{i=1}^N \frac{\hbar^2}{2m}
\sum_{\sigma,q} \int |\bnabla \varphi_i({\mbold r}, \sigma, q)|^2
d\mbold{r}\,. \label{tnph} \ee

Notice that one could define the kinetic-energy functional on the
basis of a more general set of $N$-particle wave functions:
\begin{equation}
T[n]=\inf_{\Psi\rightarrow n}<\Psi|T|\Psi>\,.
\label{Tex}
\end{equation}
However, this functional cannot be written in the form
(\ref{tnph}) and thus it is useless to derive KS equations.

Applying the variational principle to the functional $E_{HK}[n]$
with functions $\varphi_i$, $\varphi^*_i$ as functional
variables, one obtains in accordance with Eqs.~(\ref{eq2s}),
(\ref{eq4})--(\ref{tnph}) the following KS equations: \be
h_{HK}\varphi_i=\varepsilon_i\varphi_i\,, \label{ks}\ee with
\begin{equation}
h_{HK}=-\frac{\hbar^2}{2m} \Delta + U({\mbold r})\,,
\label{eq6}
\end{equation}
where $U({\mbold r}) = \delta W / \delta n$ is the local
mean-field potential and $\varepsilon_{i}$ are the Lagrange
multipliers to ensure the normalization condition of the
single-particle wave functions $\varphi_{i}$.

Often the energy functional $W[n]$ is divided into two parts:
$W[n]=E_H[n]+E_{XC}[n]$, where $E_H[n]$ is the "direct" (Hartree)
functional, while $E_{XC}[n]$ is the exchange-correlation
energy functional. Consequently, the mean field potential $U$ is
also divided into two parts. For the sake of
simplicity we will not do this in the present paper.

Eq.~(\ref{eq6}) does not contain a radial dependent effective mass
nor a spin-orbit potential which are essential ingredients of the
model nuclear single-particle Hamiltonian. To include them
we propose the following method based on a
special version of the non-local extension of the DFT. Let us
define the energy functional:
\begin{equation}
{\cal E}_0[\rho_0]=\inf_{\Psi_0\rightarrow
\rho_0}<\Psi_0|\tilde{H}|\Psi_0>\,,
\label{eq10}
\end{equation}
where $\Psi_0$ is any Slater-determinant wave function of the form
(\ref{dets}), $\rho_0$ is the single-particle DM produced by
$\Psi_0$ according to Eq.~(\ref{defrho}) (i.e. Slater-determinant
DM) and $\tilde{H}$ is an \textit{effective} many-body Hamiltonian
which generally does not coincide with the microscopic Hamiltonian
$H$.
In our approach the operator $\tilde{H}$ plays the
role of an arbitrary reference point, the choice of which will be
discussed below. We have to note that at the present moment
$\tilde{H}$ is an \textit{arbitrary} $N$-particle operator such that
the matrix element in (\ref{eq10}) exists.

The functional ${\cal E}_0[\rho_0]$ has a form of the HF energy
functional built up on the base of the effective Hamiltonian
$\tilde{H}$. So in what follows we shall refer to it also as
the HF energy functional.
Let us define the residual correlation energy $E_{RC}$ as follows
\begin{equation}
 E_{RC}[\hat{n}] = E[\hat{n}] - E_0[\hat{n}]\,,
\label{rxc}
\end{equation}
where $\hat{n}=\{n_p,n_n\}$ and
\bea
E[\hat{n}]&=&\inf_{\Psi\rightarrow\hat{n}}<\Psi|H|\Psi>\,,
\label{wn1}\\
E_0[\hat{n}]&=&\inf_{\Psi_0\to\hat{n}}<\Psi_0|\tilde{H}|\Psi_0>=\inf_{\rho_0\to\hat{n}}
\inf_{\Psi_0\to\rho_0} <\Psi_0|\tilde{H}|\Psi_0>=
\inf_{\rho_0\rightarrow\hat{n}}{\cal E}_0[\rho_0] \,. \label{eq11}
\eea The quantity $E[\hat{n}]$ is the exact energy functional
built up with the true microscopic Hamiltonian (\ref{H}) on the
set of any normalized wave functions $\Psi$. The auxiliary
functional $E_0[\hat{n}]$ (as well as the kinetic-energy
functional $T_0[n]$ in the KS theory) is defined according to the
Lieb theorem for any (not very "bad") local density $\hat{n}$. The
final energy functional of our version of the non-local DFT is
defined as:
\begin{equation}
{\cal E}[\rho_0] = {\cal E}_0[\rho_0] + E_{RC}[\hat{n}]\,,
\label{eq12}
\end{equation}
where $\rho_0$ is related to $\hat{n}$
through Eqs.(\ref{eq2s})-(\ref{defrho}).
The functionals ${\cal E}_0[\rho_0]$ and $E_{RC}[\hat{n}]$ are
defined by Eqs.~(\ref{eq10}) and (\ref{rxc}). For the moment we shall
not discuss if these functionals are known or not. The most
important for us is that they are rigorously defined.

The main property of the functional ${\cal E}[\rho_0]$
is expressed by the following equalities:
\begin{equation}
\inf_{\rho_0}{\cal E}[\rho_0] = \inf_{\hat{n}} \inf_{\rho_0 \to
\hat{n}}{\cal E}[\rho_0] = \inf_{\hat{n}}E[\hat{n}] = E_{GS}\,,
\label{eq13}
\end{equation}
where $E_{GS}$ is the \textit{true} ground state energy
of the interacting system as in the case of the HK theory.
To obtain the equations of motion we have
to suppose that the choice of the Hamiltonian $\tilde{H}$ in
Eq.~(\ref{eq10}) ensures that the infimum of the functional
${\cal E}[\rho_0]$ in (\ref{eq13}) is a minimum.
In addition, we use the general formula
for the Slater-determinant DM $\rho_0$ under the variation which
follows from Eqs.~(\ref{defrho}) and (\ref{dets}). Namely
\begin{equation}
\rho_0 (x, x') = \sum_{i=1}^N
\varphi_i(x) \varphi_i^*(x')\,,
\label{rhosl}
\end{equation}
where the sum is taken over the occupied states.  Applying the
variational principle to the functional ${\cal E}[\rho_0]$ defined
by Eq.~(\ref{eq12}) and using the functions $\varphi_i$,
$\varphi^*_i$ as functional variables according to
Eq.~(\ref{rhosl}) one will get the following set of
single-particle equations:
\begin{equation}
 \int h_0(x, x') \varphi_i (x') dx'
 + U_{RC}(x) \varphi_i (x)
 = \varepsilon_i \varphi_i (x),
\label{eq14}
\end{equation}
where we have defined the non-local Hamiltonian $h_0$  and the
local potential $U_{RC}$ as follows:
\begin{eqnarray}
 h_0(x, x') &=&
 \frac{\delta{\cal E}_0[\rho_0]}{\delta \rho_0(x', x)}\,,
\label{eq15}\\
 U_{RC}(x) &=&U_{RC}({\mbold r},q)=
 \frac{\delta E_{RC}}{\delta n_q({\mbold r})}\,.
\label{urxc}
\end{eqnarray}
It is worthwhile noting that the occupation numbers of the
Slater-determinant DM are fixed to be either 1 or 0. Thus, we avoid
 difficulties arising from the variation over the occupation
numbers that appear in the theory developed in
Ref. \cite{Gilb}.

\section{REDUCTION TO THE QUASI-LOCAL THEORY}

The approach described above enables one to introduce a reduced
energy functional ${\cal E}^{QL}_0$ which depends on the following
set of local quantities: the local particle $n_q$, kinetic-energy
$\tau_q$ and spin ${\mbold J}_q$ densities for neutrons and
protons:
\begin{eqnarray}
n_{q}({\mbold r})&=&\sum_{\sigma}\int dx'
\delta (x-x')\rho_0(x,x')\,,
\label{wn3}\\
\tau _{q}({\mbold r})&=&\sum_{\sigma}\int dx'
\delta (x-x')({\bnabla}_{r}{\bnabla}_{r'})\rho_0(x,x')\,,
\label{wn4}\\
{\mbold J}_{q}({\mbold r})&=&i\sum_{\sigma}\int dx' \delta({\mbold
r}-{\mbold r}') \delta_{q,q'} [({\mbold\sigma})_{\sigma',\sigma}
\times\bnabla_{r}] \rho_0(x,x')\,, \label{eq16}
\end{eqnarray}
where $\delta (x-x') = \delta({\mbold r}-{\mbold r}')
\delta_{\sigma,\sigma'} \delta_{q,q'}$, the quantities $\tau_{q}$
and ${\mbold J}_{q}$ are the \textit{uncorrelated} neutron and
proton kinetic-energy and spin densities respectively. Introducing
the short notation $\rho_{QL}\equiv\{n_p,n_n,\tau_p,\tau_n,{\mbold
J}_p,{\mbold J}_n\}$, let us define the quasi-local energy
functional as follows:
\be {\cal E}^{QL}[\rho_{QL}]={\cal
E}^{QL}_0[\rho_{QL}]+E_{RC}[\hat{n}]\,,
\label{QLF}
\ee
where
\be
{\cal E}^{QL}_0[\rho_{QL}]=
\inf_{\rho_0\rightarrow \rho_{QL}}{\cal E}_0[\rho_0]\,.
\label{QLF0}
\ee
Notice, that the many-to-one mapping
$\rho_{0}\to\rho_{QL}$ is established according to
Eqs.~(\ref{wn3})-(\ref{eq16}), and
that the set $\hat{n}=\{n_p,n_n\}$ enters $\rho_{QL}$:
$\hat{n}\in \rho_{QL}$.

From Eqs.~(\ref{eq12}), (\ref{eq13}), (\ref{QLF}) and (\ref{QLF0})
we have
\be
\inf_{\rho_{QL}} {\cal E}^{QL}[\rho_{QL}] = E_{GS}\,.
\ee
Using Eq.~(\ref{eq4a}) and the explicit
expressions for the remaining local quantities: \bea \tau_q({\mbold
r})&=&\sum_{i=1}^N \sum_{\sigma}|\bnabla\varphi_i ({\mbold
r},\sigma,q)|^2\,,
\label{wn5}\\
{\mbold J}_q({\mbold r})&=& i \sum_{i=1}^N \sum_{\sigma,\sigma'}
\varphi_i^*({\mbold r},\sigma',q) [({\mbold
\sigma})_{\sigma',\sigma}\times{\bnabla}]\varphi_i({\mbold
r},\sigma,q)\,, \label{tJ} \eea one can apply the variational
principle to  the energy functional ${\cal E}^{QL}[\rho_{QL}]$
with $\varphi_i$, $\varphi_i^*$ as functional variables. The
resulting single-particle equations are: \be
h_q\varphi_i=\varepsilon_i\varphi_i\,, \label{QLSP}\ee
where
\be
h_q=-\bnabla\frac{\hbar^2}{2m_q^*({\mbold r})}\bnabla+U_q({\mbold r})
- i {\mbold W}_q({\mbold r}) \cdot [\bnabla\times{\mbold\sigma}]\,,
\ee
and
\bea
\frac{\hbar^2}{2m_q^*({\mbold r})}&=&\frac{\delta{\cal
E}^{QL}}{\delta\tau_q({\mbold r})}\,,
\label{wn6}\\
U_q({\mbold r})&=&\frac{\delta{\cal E}^{QL}}
{\delta n_q({\mbold r})}\,,
\label{wn7}\\
{\mbold W}_q ({\mbold r})&=&\frac{\delta{\cal E}^{QL}}
{\delta{\mbold J}_q({\mbold r})}\,.
\label{SPP}\eea
If the functional ${\cal E}^{QL}$ were known,
one would be able to calculate the exact ground state
energy $E_0 = E_{GS}$ and exact local densities $n_q = n_{q, GS}$.
The kinetic-energy density $\tau_q$ and spin density ${\mbold J}_q$
correspond to the system without correlations and
do not coincide with the exact densities in our approach.

The following remark is in order. As was pointed in the
Introduction, the energy functional dependence on the exact
kinetic-energy and spin densities could be introduced, for example,
by the above procedure of the quasi-local reduction
being applied to the energy functional of the non-local theory
developed in Ref.~\cite{Gilb}. However, the mapping of the
Slater-determinant wave functions $\Psi_0$ onto the \textit{exact}
kinetic-energy and spin densities is not defined, as opposed to
the mapping onto the \textit{exact} local density, i.e.
Eqs.~(\ref{wn5}) and (\ref{tJ}) are wrong for the exact densities.
As a consequence, it is impossible to derive the single-particle
equations such as (\ref{QLSP}). The use of the Slater-determinant DM
($\rho_0$ in our approach) enables us to avoid this problem.

\section{EXTENDED THOMAS-FERMI APPROXIMATION \protect\\
         IN THE QUASI-LOCAL DFT}

Now we are turn to the discussion of the choice of the effective
operator $\tilde{H}$. First, the energy functional ${\cal E}_0$ has
to be well defined. While the functional $E[\hat{n}]$ is always
defined, this is not the case for the functionals ${\cal
E}_0[\rho_0]$ and $E_0[\hat{n}]$. Indeed, the matrix element of the
microscopic Hamiltonian with the bare nucleon-nucleon (NN)
interactions over the Slater-determinant wave functions can
diverge due to the short-range singularity of the force as it is
happens in the standard many-body theory. Thus, we will use a
Hamiltonian $\tilde{H}$ with an \textit{effective} NN interaction
whose matrix elements are well defined. Second, the local energy
functional $E[\hat{n}]$ obviously has the minimal property that is
necessary to apply the variational principle; however, it can be not true
for the non-local energy functional ${\cal E}[\rho_0]$ (in spite
of the fact that the equality (\ref{eq13}) is always true).
Hence one has to choose $\tilde{H}$ to ensure the
minimal property of the energy functional ${\cal E}[\rho_0]$.
Notice that at least in one particular case, when $\tilde{H} = T$,
this condition is fulfilled because we come to the usual KS theory.
Indeed, if one sets $\tilde{H}=T$ then $m_q^* = m$,
${\mbold W}_q = 0$, and Eq.~(\ref{QLSP}) coincides with the
KS equation. In this case the residual correlation energy
functional $E_{RC}[\hat{n}]$ corresponds to the
sum of the direct (Hartree) term and the
exchange and correlation energy in the KS method.

Here we consider a more general case, setting
$\tilde{H}$ to be the $N$-particle Hamiltonian with an
\textit{effective} NN interaction: \be \tilde{H}=T+\sum_{i\ne
j}\tilde{v}^{NN}_{ij}+\sum_{i\ne j}v^{Coul}_{ij}.
 \label{HT}\ee
We will use the effective NN force of the form \be
\tilde{v}^{NN}_{ij}=\hat{v}^c_{ij}+\hat{v}^{so}_{ij},
\label{vef}\ee
where the central part of the effective force is given by
\begin{equation}
\hat{v}^c_{ij}=\sum_n [w_n + b_n P^{\sigma}_{ij} -
h_n P^{\tau}_{ij} - m_n P^{\sigma}_{ij} P^{\tau}_{ij}]
v_n(s), \label{NN}
\end{equation}
$w_n,b_n,h_n,m_n$ are the parameters of the force
($n=1,2,\ldots$), $P^{\sigma}_{ij}$ and $P^{\tau}_{ij}$
are the spin and isospin exchange operators and $v_n (s)$
(${\mbox{\boldmath$s$}}={\mbox{\boldmath$r$}}_i -
{\mbox{\boldmath$r$}}_j$)
are the radial formfactors of the central part of the effective
force. For the sake of simplicity we will consider only one term in
the sum (\ref{NN}) i.e. we will drop index $n$ in the following.
The spin-orbit part of the force is chosen as follows
\be
\hat{v}^{so}_{ij}=iW_0({\mbold \sigma}_i+{\mbold
\sigma}_j) \cdot [{\mbold k}'\times\delta
({\mbold r}_i - {\mbold r}_j){\mbold k}]\,,
\label{vso}
\ee
where
${\mbold k}=\frac{1}{2i}(\bnabla_i - \bnabla_j)$ denotes the
operator acting on the right and
${\mbold k}'=-\frac{1}{2i}(\bnabla_i-\bnabla_j)$ is the
operator acting on the left.

It was mentioned that the choice of the effective Hamiltonian
$\tilde{H}$ and the effective interactions is rather arbitrary.
Practically the operator $\tilde{H}$ is confined only by the
above formal mathematical conditions. The situation is
quite different in the usual HF method where there are no other
ingredients apart from the effective forces which are taken to be
density dependent in order to ensure the nuclear saturation. In
our approach the particular choice of $\tilde{H}$ is
compensated by the addition of the residual correlation energy
$E_{RC}[\hat{n}]$, which contains all necessary density dependence
of the total energy functional. In the applications of the method
the functional $E_{RC}[\hat{n}]$ is parametrized
phenomenologically, the parameters are adjusted to describe
nuclear ground-state properties. Following this ideology the
effective interactions entering operator $\tilde{H}$ are taken to
be density independent. Thus, we would like to stress
that the effective interactions in our DFT approach are
not exactly the same as in HF theory.

Let us define the density ${\cal H}_0$ of the quasi-local HF
energy functional ${\cal E}_0^{QL} [\rho_{QL}]$ as follows
\be
{\cal E}_0^{QL}=\int d{\mbold r}{\cal H}_0({\mbold r})\,.
\label{e0ql}
\ee
According to (\ref{HT})-(\ref{vso})
the energy density ${\cal H}_0$ is described by six terms:
\be
{\cal H}_0 = \frac{\hbar^2}{2m} (\tau_n + \tau_p) + {\cal
H}^{Nucl}_{Dir}+{\cal H}^{Nucl}_{Exch}+{\cal H}^{Coul}_{Dir}+{\cal
H}^{Coul}_{Exch}+{\cal H}^{so}\,.
\label{five}\ee
The direct nuclear energy density ${\cal H}^{Nucl}_{Dir}$ comes from the
central part of the NN force and is given by
\be
{\cal H}^{Nucl}_{Dir}({\mbold r}) =\frac{1}{2}\int d{\mbold
r}'\{(w+\frac{b}{2})n({\mbold r})n({\mbold
r}')-(h+\frac{m}{2})[n_p({\mbold r})n_p({\mbold r}')+n_n({\mbold
r})n_n({\mbold r}')] \} v(|{\mbold r} - {\mbold r}'|)\,.
\label{hnd} \ee
The density of the Coulomb direct energy is
\begin{equation}
{\cal H} ^{Coul}_{Dir}({\mbox{\boldmath$r$}}) =
\frac{e^{2}}{2} \int d{\mbox{\boldmath$r$}}%
^{\prime }\frac{n_{p}(\mbox{\boldmath$r$})
n_{p}({\mbox{\boldmath$r$}}^{\prime })} {|{\mbox{\boldmath$r$}} -
{\mbox{\boldmath$r$}}^{\prime }|}.
\label{eq25a} \end{equation}
These direct energies give the contribution to the so-called
Hartree functional. To calculate the exchange terms that come from
the central part of the NN force we use the recently proposed ETF
approximation for the DM up to $\hbar^2$ order \cite{SV}.
Notice that there are other possible options
to obtain the quasi-local energy
functional based on the Negele-Vautherin and
Campi-Bouyssy DM expansions \cite{Hof,CB,NV}.
In our approach
for spin-saturated nuclei the nuclear exchange energy density is
given by two terms \be {\cal H}^{Nucl}_{Exch}={\cal
H}^{Nucl}_{Exch,0}+{\cal H}^{Nucl}_{Exch,2}\,.\label{HNE} \ee
The first term is calculated to $\hbar^0$ order
(which corresponds to the Slater approximation for the DM):
\begin{equation}
{\cal H}^{Nucl}_{Exch,0}({\mbold r}) =
\int d{\mbold s} v(s) \left( \frac{1}{2}
X_{e1} \sum_q \left(n_q({\mbold r})
\hat{j}_1(k_q s)\right)^2
+ X_{e2} n_n({\mbold r}) \hat{j}_1(k_n s)
         n_p({\mbold r}) \hat{j}_1(k_p s) \right),
\label{HNE0}
\end{equation}
where $k_q({\mbold r})=(3\pi^2n_q({\mbold r}))^{1/3}$ is the Fermi
momentum, $\hat{j}_1(x) = 3 j_1(x) / x$,
$j_1(x)$ is the spherical Bessel function and
$X_{e1}=m+h/2-b-w/2$, $X_{e2}=m+h/2$. The second term corresponds
to the $\hbar^2$ correction:
\begin{equation}
{\cal H}^{Nucl}_{Exch,2}({\mbold r})=\sum_q\frac{\hbar ^{2}}{2m}\big[%
(f_{q}-1)(\tau _{q}-\frac{3}{5}k_{q}^{2}n _{q}-\frac{1}{4}\Delta
n_{q})+k_{q} f'_{q}
(\frac{1}{27}\frac{(\bnabla n_q)^2}{n_q}
-\frac{1}{36}\Delta n _{q})\big].
\label{HNE2} \end{equation}
In this equation $f_{q} = f_{q}(\mbox{\boldmath$r$},k_q)$,
$f'_{q} = (\partial f_{q}(\mbox{\boldmath$r$},k) / \partial k)_
{k=k_q}$.
The function $f_q(\mbox{\boldmath$r$},k)$ is
the inverse of the position- and momentum-dependent effective mass
and is defined in the ETF approximation by
\begin{equation}
f_{q}(\mbox{\boldmath$r$},k)
=1+\frac{m}{\hbar ^{2}k}\frac{\partial
V^{Nucl}_{Exch,q}(\mbox{\boldmath$r$},k)}{\partial k},
\label{eq25b}
\end{equation}
where $V^{Nucl}_{Exch,q}$ is the Wigner transform of the
exchange potential in the Thomas-Fermi approximation:
\be V^{Nucl}_{Exch,p}({\mbold r},k)=
\int d{\mbold s} e^{-i{\mbold k}{\mbold s}} v(s)
[X_{e1} n_p({\mbold r}) \hat{j}_1(k_p s)
+X_{e2} n_n({\mbold r}) \hat{j}_1(k_n s)]
\label{eq25c} \ee
and analogously for $V^{Nucl}_{Exch,n}$
with the permutation of indices $p$ and $n$
(see \cite{SV} for details).
It is worthwhile noting that within
the semiclassical ETF approximation the kinetic-energy density is a
functional of the local density. Then the energy functional obtained
would only depend on the local particle density and spin density.
However, it
was found in \cite{SV} that the use of the quantal kinetic energy
in Eq.(\ref{HNE2}), which yields the radial-dependent effective mass,
improves agreement with results of the full HF
calculation significantly. It motivates us to use the anzatz
(\ref{wn5}) for $\tau_q$ in the present paper.

The Coulomb exchange energy consists of the Slater term and the
second-order correction that in the ETF approximation is written as
\cite{DG,SV}:
\begin{equation}
{\cal H}^{Coul}_{Exch}({\mbox{\boldmath$r$}})=-\frac{3}{4}
\left( \frac{3}{\pi} \right)^{1/3}  n_{p}^{4/3}
- \frac{7}{432\pi (3\pi ^{2})^{1/3}}
\frac{(\bnabla n_{p})^{2}}{n_{p}^{4/3}}.
\label{exc}
\end{equation}
Finally, the spin-orbit energy density is given by
\be {\cal H}^{so}({\mbold r})=-\frac{1}{2}W_0[n({\mbold r})
\bnabla{\mbold J}+n_n({\mbold r})\bnabla{\mbold J}_n
+n_p({\mbold r})\bnabla{\mbold J}_p]\,,
\label{hso}
\ee
where ${\mbold J} = {\mbold J}_p + {\mbold J}_n$.

It is worthwhile noting that in this
section we replace the exact quasi-local functional ${\cal
E}_0^{QL}$ by the approximate functional calculated within the ETF
approximation. The difference between them gives a very small
contribution (see the following sections), but it cannot be
totally included within the residual correlation term because it depends
on the $\rho_{QL}$ while the latter only depends on the $\hat{n}$.
The determination of the residual correlation energy $E_{RC}[\hat{n}]$
will be specified in the next section.

\section{THE ENERGY FUNCTIONAL AND SINGLE-PARTICLE \protect\\
         EQUATIONS FOR THE GOGNY FORCE}

     The formulae (\ref{hnd}), (\ref{HNE})-(\ref{eq25c})
are valid for any radial formfactor $v(s)$ of the central part
of the effective forces. Here we give the explicit expressions
for the above-defined quantities in the case of a
Gaussian formfactor $v(s)=exp(-s^2/a^2)$ entering in the Gogny-type
effective forces which are used in the numerical applications
of our method. Assuming spherical symmetry of the particle densities
we get from Eqs.~(\ref{hnd}) and (\ref{HNE0})
\bea
{\cal H}^{Nucl}_{Dir} ({\mbold r}) &=&
\frac{\pi a^2}{2r}\int_0^{\infty}dr' r'
\{\exp \left( -\frac{(r-r')^2}{a^2} \right)
- \exp \left( -\frac{(r+r')^2}{a^2} \right)\}\nonumber\\
&\times&\{(w+\frac{b}{2})n(r)n(r')
-(h+\frac{m}{2})[n_p(r)n_p(r')+n_n(r)n_n(r')]\}\,,
\label{EDN}
\eea
\bea
{\cal H}^{Nucl}_{Exch,0} ({\mbold r}) &=&
\frac{2}{3\pi^{5/2}a^3}
\nonumber\\
&\times&\{ X_{e1} \sum_q \left( \frac{\sqrt{\pi}}{2}a^3k_q^3
\mbox{erf}(ak_q)+\left( \frac{a^2k_q^2}{2}-1 \right)
\exp(-a^2k_q^2) - \frac{3a^2k_q^2}{2}+1 \right)
\nonumber\\
&+& X_{e2}
\sum_{\eta=\pm 1}\eta
[ \frac{\sqrt{\pi}}{2}a^3(k_{n}+\eta
k_p)(k_n^2+k_p^2-\eta k_n k_p)
\mbox{erf} \left( \frac{a}{2}(k_n+\eta k_p) \right)
\nonumber\\
&+& \left( a^2 (k_n^2+k_p^2-\eta k_n k_p) - 2 \right)
\exp \left(-\frac{a^2}{4}(k_n+\eta k_p)^2 \right) ] \}\,.
\label{HNE0b}
\eea
The second-order correction to the exchange nuclear energy
density (\ref{HNE2}) can be rewritten in the following way:
\be {\cal H}^{Nucl}_{Exch,2} ({\mbold r}) =
\sum_q \left( F_q(\tau_q-\frac{3}{5}k_q^2n_q-\frac{1}{4}\Delta n_q)
+G_q (\frac{1}{27}\frac{(\bnabla n_q)^2}{n_q}- \frac{1}{36}\Delta
n_q)\right)\,,
\label{eq29}
\ee
where the explicit value of the functions $F_q$ and $G_q$  calculated with
a Gaussian formfactor are given in Appendix 1.

The SPP is defined following Eq.(\ref{wn7}).
According to (\ref{QLF}), (\ref{e0ql})
and (\ref{five}) we split it into five pieces:
 \be
U_q=U^{Nucl}_{Dir,q}+U^{Nucl}_{Exch,q}+U^{Coul}_{Dir,q}
+U^{Coul}_{Exch,q}+U^{RC}_{q},
\label{UG}\ee
where the direct nuclear SPP is given by
\bea
U^{Nucl}_{Dir,q}({\mbold r}) &=&
\frac{\pi a^2}{r}\int_0^{\infty}dr' r'
\{\exp \left( -\frac{(r-r')^2}{a^2} \right)
- \exp \left( -\frac{(r+r')^2}{a^2} \right)\}\nonumber\\
&\times&
[(w+\frac{b}{2})n(r')-(h+\frac{m}{2})n_q(r')].
\label{UND}
\eea
The exchange nuclear potential consists of two parts following
(\ref{HNE}):
\be
U^{Nucl}_{Exch,q}=U^{Nucl}_{Exch,q,0}+U^{Nucl}_{Exch,q,2}\;,
\ee
where, for example, the Slater part of the exchange SPP
acting on the protons is given by
\bea
U^{Nucl}_{Exch,p,0}&=&\frac{2}{\sqrt{\pi}a^3k_p^3}\{X_{e1}
[\frac{\sqrt{\pi}}{2}a^3k_p^3
\mbox{erf}(ak_p)+a^2k_p^2 \exp(-a^2k_p^2)-a^2k_p^2]
\nonumber\\
&+&2X_{e2} \sum_{\eta = \pm 1}\eta 
[\frac{\sqrt{\pi}}{4}a^3 k_p^3
\mbox{erf} (\frac{a}{2}(k_p+\eta k_n))
+\frac{1}{2}a^2 k_p^2 
\exp (-\frac{a^2}{4}(k_p+\eta k_n)^2 ) ]\}.
\label{eq31b}
\eea
For the second-order contribution to the exchange SPP, we have
\bea U^{Nucl}_{Exch,p,2}&&=\pi^2\{\frac{1}{k_p}[
F^p_p(\tau_p-\frac{3}{5}k_p^2 n_p)+\frac{1}{27}
\left( \frac{3G_p}{k_p^2}-G^p_p \right)
\frac{(\bnabla n_p)^2}{n_p}
\nonumber\\
&&-\frac{1}{36} \left( \frac{8G_p}{k_p^2}+G^p_p+9F^p_p \right)
\Delta n_p + F^p_n(\tau_{n}-\frac{3}{5}k_{n}^2 n_{n})
+\frac{1}{27} G^p_n\frac{(\bnabla n_{n})^2}{n_{n}}
\nonumber\\
&&-\frac{1}{36}(G^p_n+9F^p_n) \Delta n_{n}]
-\frac{2}{27k_n}G^n_p\frac{(\bnabla n_p)(\bnabla n_{n})}
{n_p}\}\nonumber\\&&-F_pk_p^2-\frac{1}{4}\Delta
F_p-\frac{1}{36}\Delta G_p, \label{eq32} \eea
where the functions $F_q^{q'}$, $G_q^{q'}$,
$\Delta F_q$ and $\Delta G_q$ used in Eq.~(\ref{eq32})
are also given in Appendix 1.
The formulae for the Slater and $\hbar^2$ contributions to the nuclear
exchange potential acting on neutrons are obtained by replacing $n$ by $p$
and $p$ by $n$ in Eqs.(\ref{eq31b}) and (\ref{eq32}).

The Coulomb direct and exchange potentials, entering
Eq.~(\ref{UG}), are not equal to zero only for protons.
In the explicit form we have
\bea
U^{Coul}_{Dir,p}({\mbold r}) &=& e^2 \int d{\mbold r}'
\frac{n_p({\mbold r}')}{|{\mbold r} - {\mbold r}'|}\,,\\
U^{Coul}_{Exch,p}({\mbold r}) &=&
- \left( \frac{3}{\pi} n_p ({\mbold r}) \right)^{1/3}\,.
\eea
The inclusion of the
$\hbar^2$-correction to the Coulomb exchange energy
(second term in Eq.~(\ref{exc})) in the SPP
leads to the unphysical behaviour of the potential
at $r \to \infty$. We thus only calculate its contribution to
the binding energy as a perturbation.

The radial dependent effective mass $m^*_q({\mbold r})$ and
the form factor ${\mbold W}_q ({\mbold r})$ of the spin-orbit
potential are defined according to Eqs.~(\ref{wn6}), (\ref{SPP}),
(\ref{e0ql}), (\ref{five}), (\ref{hso}) and (\ref{eq29})
by the relations
\bea
\frac{\hbar^2}{2 m^*_q ({\mbold r})} &=&
\frac{\hbar^2}{2 m} + F_q\,,
\label{efm}\\
{\mbold W}_q ({\mbold r}) &=&
\frac{1}{2} W_0 (\bnabla n + \bnabla n_q )\,.
\eea

Now we shall turn to the determination of the
residual correlation energy.
We take it in the form of the phenomenological ansatz
\be E_{RC}[\hat{n}]= \frac{t_3}{4} \int d{\mbold r}
n^{\alpha}({\mbold r}) [(2+x_3) n^2({\mbold r}) - (2 x_3 + 1)
\left( n_p^2({\mbold r}) + n_n^2({\mbold r}) \right)].
\label{ERXC}\ee
The parameters $t_3$, $x_3$ and $\alpha$ together with
the parameters of the Hamiltonian $\tilde {H}$ have to be
chosen from the condition of the best description of the
nuclear ground-state properties. In the calculations, which
are presented in the next section, we use the well-known parameter
set of the D1S Gogny force. Let us note that the formula (\ref{ERXC})
is the standard ansatz which enters not only a
density-dependent part of the Gogny forces but, for example,
a density-dependent part of the Skyrme forces.
Eq.~(\ref{ERXC}) leads to the following contribution into
the SPP in accordance with the definition (\ref{UG}):
\be
U^{RC}_q = \frac{t_3}{4} n^{\alpha - 1}
\left[ ( 2 + \alpha ) (2+x_3) n^2
- (2 x_3 + 1)
\left( \alpha (n_p^2 + n_n^2) + 2 n_q n \right) \right]\,.
\ee

\section{NUMERICAL RESULTS}

In this section we want to check our local approximation to the HF
method using finite-range Gogny forces. First of all, we compare
the exact HF ground-state binding energies as well as the rms
radii of the neutron and proton densities of some magic nuclei
computed with the Gogny D1S force \cite{BGG} with our DFT results.
In this comparison we use two different quasi-local functionals :
DFT-$\hbar^0$, where the exchange energy coming from the
finite-range part of the interaction is taken at pure Thomas-Fermi
level (Slater approach) and DFT-$\hbar^2$ where the ETF-$\hbar^2$
contributions have been added to the Slater part. Notice that in
this DFT-$\hbar^2$ approach the semiclassical kinetic-energy
density entering Eq.~(\ref{HNE2}) has been replaced by the
corresponding quantal density for the reasons pointed out above.
In both DFT calculations we solve the local Schr\"odinger equation
(\ref{QLSP}) for neutrons and protons with the potentials and
effective masses reported in the previous section. Table~\ref{tab1}
collects all these binding energies and radii which have been
computed taking into account the two-body centre-of-mass
correction. In our calculation we take into account this
correction as explained in Appendix 2. We would like to say in
passing that the numerical value of this two-body centre-of-mass
correction along the whole periodic table is very well reproduced
by using the pocket formula based on the harmonic oscillator and
derived in \cite{Mart}.

From Table~\ref{tab1} we can see that the DFT-$\hbar^2$
binding energies reproduce the HF values fairly well.
The differences between HF and DFT-$\hbar^2$
are smaller than 1\% for all the considered nuclei from $^{40}$Ca to
$^{208}$Pb and in the case of $^{16}$O the relative difference is only
1.8\%. The DFT-$\hbar^0$ binding energies show larger discrepancies
with the full HF results. The relative differences range from 7\% in
$^{16}$O to 1\% in the heaviest nucleus considered $^{208}$Pb. As regards
the rms radii of the neutron and proton densities, the full HF values are
again better reproduced by the DFT-$\hbar^2$ approximation than by the
DFT-$\hbar^0$ approach. These results show the importance of the
$\hbar^2$ corrections in the local approximation to the HF exchange
energy.
It should be pointed out that the eigenvalues $\varepsilon_i$
in Eq.~(\ref{QLSP}) have no rigorous physical sense in the DFT
except for the energy of the last filled level, which corresponds
to the neutron or proton separation energy (chemical potential).
Table~\ref{tab1} also displays the neutron and proton
chemical potentials obtained using the DFT-$\hbar^0$, DFT-$\hbar^2$
and HF approximations. The DFT-$\hbar^2$ chemical potentials differ
from the HF chemical potentials by less than 1 MeV while the shift of
the DFT-$\hbar^0$ separation energies with respect to the the full HF
values is larger and can be around 3 MeV for light nuclei.

\begin{table}
\begin{center}
\caption{\label{tab1}
Total binding energies $B$ (in MeV), neutron $r_n$ and proton
$r_p$ rms radii (in fm) and separation energies of neutrons $S_n$
and protons $S_p$ (in MeV) of some magic nuclei computed with the
D1S Gogny force using the DFT-$\hbar^0$ and DFT-$\hbar^2$
approaches compared with the full HF results.}
\vspace{1em}
\begin{tabular}{l l c c c c c c c c}
\hline
\hline
&  $\vphantom{\frac{R^2}{2}}$
&  $^{16}$O  &  $^{40}$Ca &  $^{48}$Ca &  $^{90}$Zr
   & $^{132}$Sn & $^{208}$Pb \\
\hline
$B$: & DFT-$\hbar^0$ &
120.2 & 329.6 & 407.5 & 772.1 & 1092.9 & 1623.3
\\
& DFT-$\hbar^2$ &
127.3 & 341.9 & 415.0 & 783.9 & 1101.2 & 1636.6
\\
& HF &
129.6 & 344.6 & 416.7 & 785.6 & 1103.0 & 1638.9
\\
& exp &
127.6 & 342.1 & 416.0 & 783.9 & 1102.9 & 1636.4
\\
$r_{\rm n}$:
& DFT-$\hbar^0$ &
2.72 & 3.41 & 3.62 & 4.29 & 4.87 & 5.59  \\
& DFT-$\hbar^2$ &
2.69 & 3.39 & 3.61 & 4.28 & 4.86 & 5.58  \\
& HF &
2.65 & 3.37 & 3.58 & 4.27 & 4.84 & 5.57 \\
$r_{\rm p}$:
& DFT-$\hbar^0$ &
2.75 & 3.46 & 3.47 & 4.24 & 4.66 & 5.44 \\
& DFT-$\hbar^2$ &
2.71 & 3.44 & 3.46 & 4.23 & 4.66 & 5.44 \\
& HF  &
2.67 & 3.41 & 3.44 & 4.21 & 4.65 & 5.44 \\
$S_{\rm n}$:
& DFT-$\hbar^0$ &
12.20 & 13.21 &  9.31 & 11.87 &  7.49 &  7.45 \\
& DFT-$\hbar^2$ &
14.55 & 15.36 &  9.52 & 12.02 &  7.59 &  8.03  \\
& HF &
15.08 & 16.04 &  9.66 & 11.88 &  7.68 &  7.80 \\
& exp &
15.66 & 15.64 &  9.95 & 11.97 &  7.31 &  7.37 \\
$S_{\rm p}$:
& DFT-$\hbar^0$ &
 8.98 &  6.43 & 14.07 &  7.43 & 15.52 &  8.17  \\
& DFT-$\hbar^2$ &
11.24 &  8.45 & 16.51 &  8.25 & 15.94 &  9.29  \\
& HF &
12.53 &  9.27 & 17.09 &  8.36 & 16.23 &  9.51  \\
& exp &
12.13 &  8.33 & 15.81 &  8.36 & 15.34 &  8.01  \\
\hline
\hline
\end{tabular}
\end{center}
\end{table}

It is important to note that the agreement of the proposed DFT
approximations with the full HF results is determined by the
treatment of the non-local effects.
The contribution of the pointed effects
can be quantified in terms of the effective mass in nuclear matter.
The results of Table~\ref{tab1} show that for effective forces with
an effective mass over the bare mass of around 0.7, as in the case of
the Gogny forces \cite{BBDG}, the non-local effects can be
very well accounted for the DFT-$\hbar^2$ functional proposed in
this paper. However, for forces where the non-local effects are
larger, the result of our DFT-$\hbar^2$ approximation is worse
when compared with the HF results, but is still  better than other
suitable choices of the exchange-energy localization such as the
Negele-Vautherin or Campi-Bouyssy approaches (see Ref.\cite{SV}
for more details about this point).

\begin{figure}
\includegraphics*[scale=0.9]{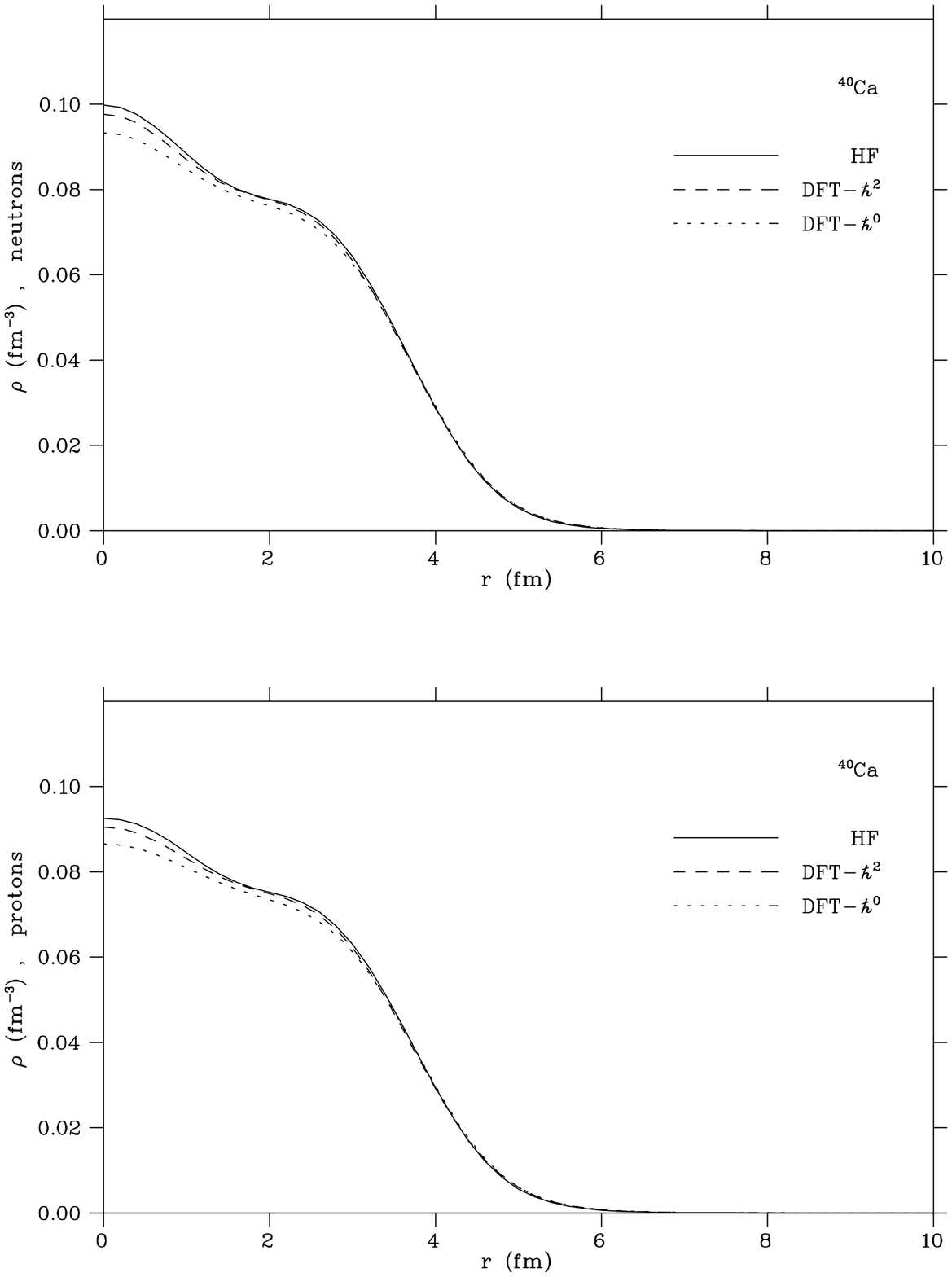}
\caption{\label{fig1}
Neutron and proton densities of the nucleus $^{40}$Ca calculated
with the D1S Gogny force using the DFT-$\hbar^0$ (short-dashed line)
and DFT-$\hbar^2$ (dashed line) approaches compared with the full
HF densities (solid line).}
\end{figure}

\begin{figure}
\includegraphics*[scale=0.9]{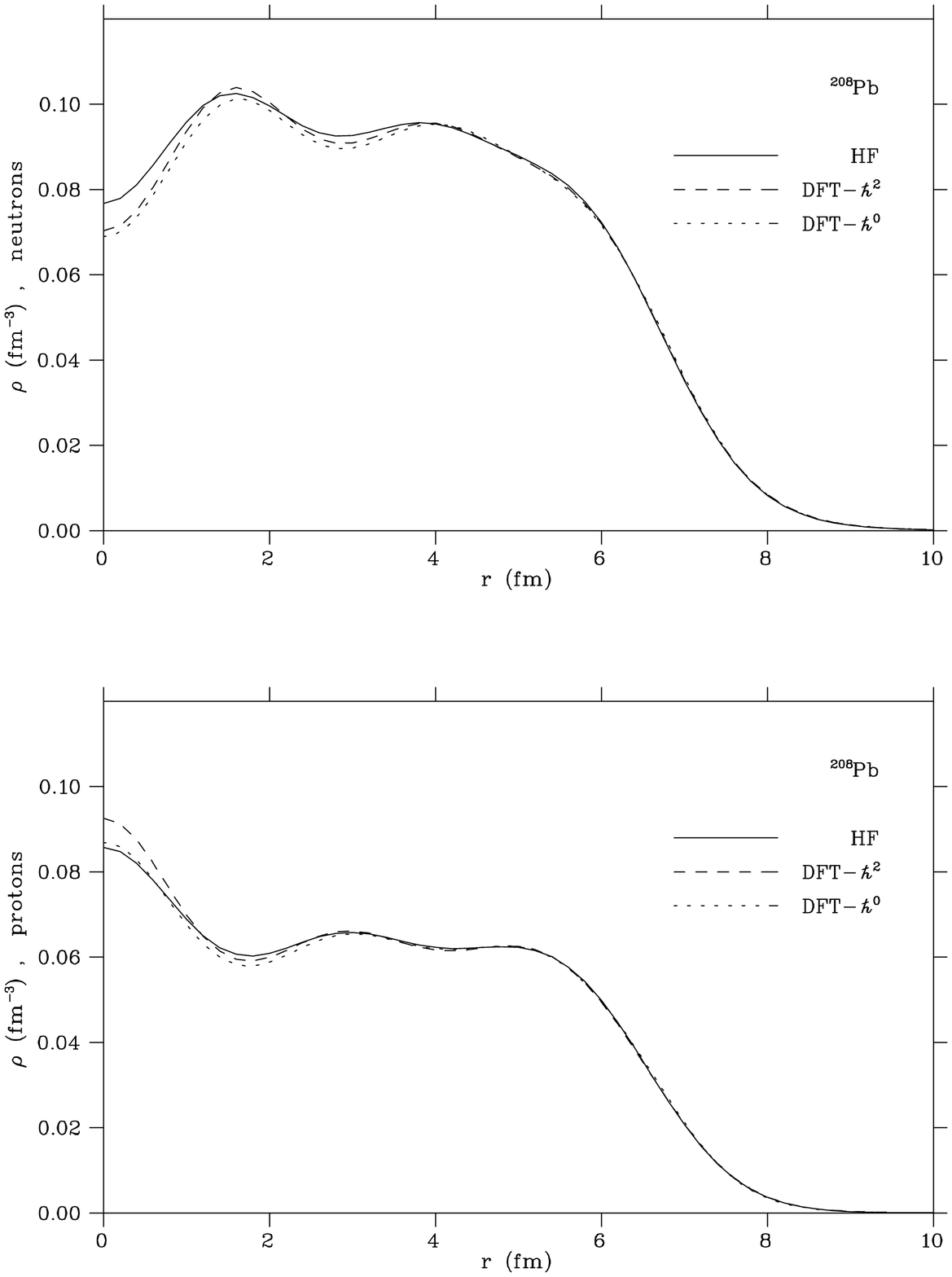}
\caption{\label{fig2}
Same as Figure \ref{fig1} for the nucleus $^{208}$Pb.}
\end{figure}

Figures \ref{fig1} and \ref{fig2} displays the neutron and proton
densities for $^{40}$Ca and $^{208}$Pb obtained using
the D1S force in the full HF (solid line)
as well as in the DFT-$\hbar^2$ (dashed line) and
DFT-$\hbar^0$ (short-dashed line) calculations.
The DFT proton and neutron densities
nicely reproduce the surface and the tail of the
full HF densities. In the region near the centre of the nuclei,
the DFT density follows the full HF density profile very well
although a small shift between the full HF and DFT proton and
neutron densities appears in this central region. These
differences can be attributed to the fact that our DFT description
of the $s$-orbitals, whose wave functions mainly provide the proton
and neutron densities at the centre of the nuclei, show some small
differences with the corresponding HF $s$-orbitals. Comparing the
DFT-$\hbar^0$ and DFT-$\hbar^2$ densities, it can be seen that
including the $\hbar^2$ contributions in our local approximation,
one obtains a better agreement with the full HF densities.

Figures 3 and 4 display the radial dependence of the neutron and
proton effective masses over the bare nucleon mass calculated with
the DFT-$\hbar^2$ approach (see Eq. (\ref{efm})) for $^{40}$Ca and
$^{208}$Pb nuclei (solid lines).
\begin{figure}
\includegraphics*[scale=0.9]{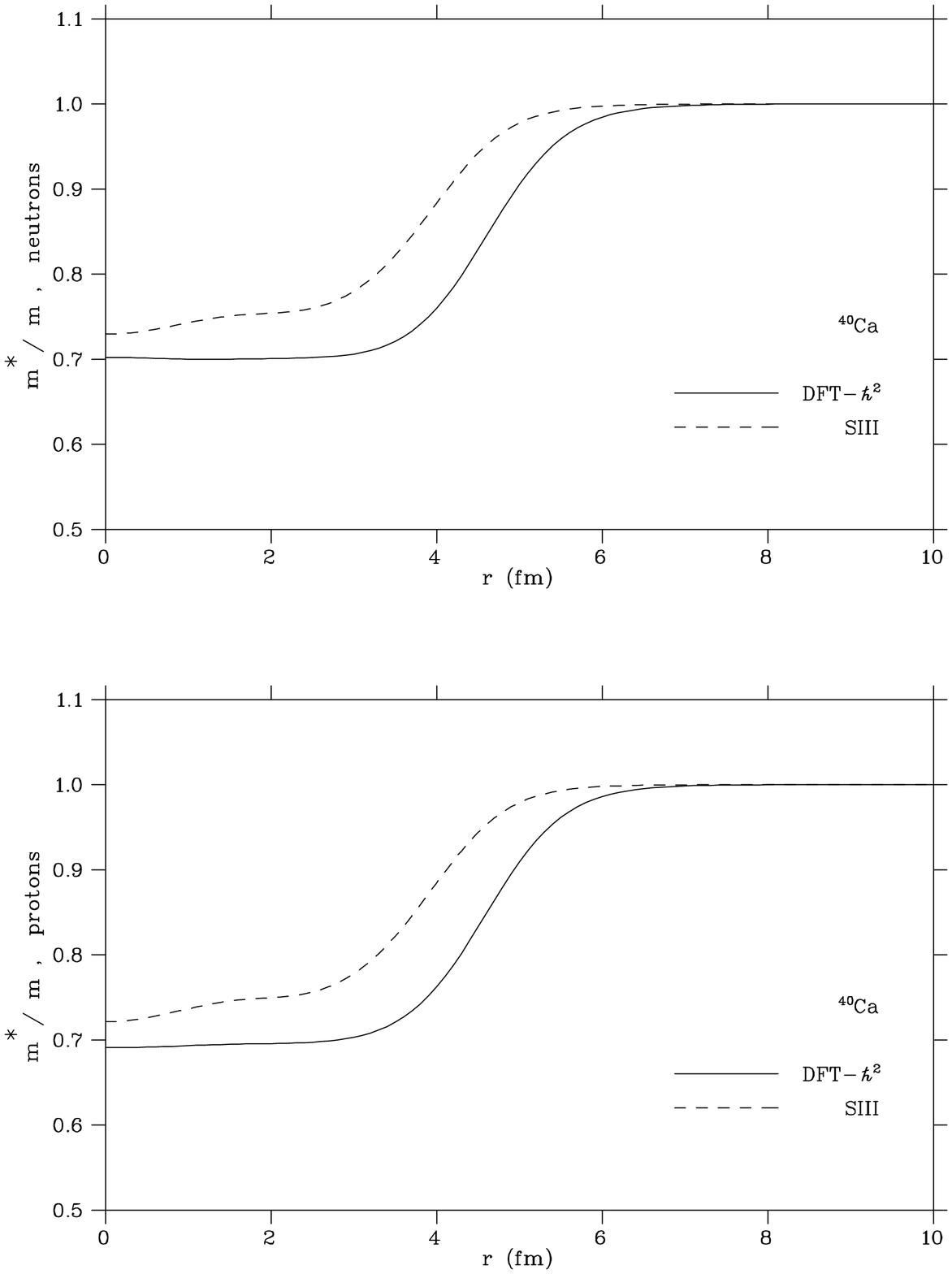}
\caption{\label{fig3}
Neutron and proton radial dependence of the effective mass of
the nucleus $^{40}$Ca calculated with the D1S Gogny force using the
DFT-$\hbar^2$ approach (solid line) compared with the corresponding HF
effective masses obtained with the Skyrme SIII force (dashed line).}
\end{figure}
\begin{figure}
\includegraphics*[scale=0.9]{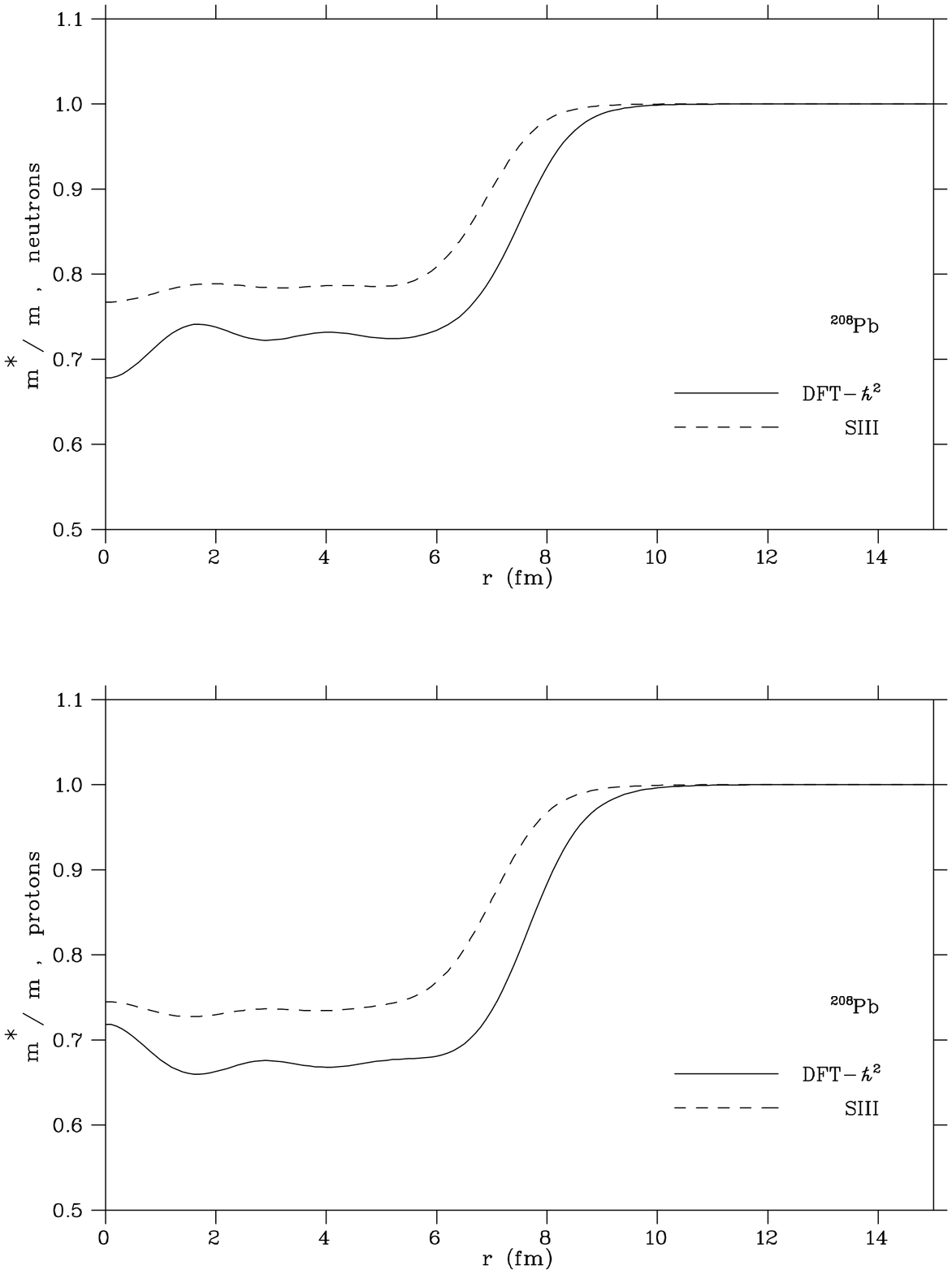}
\caption{\label{fig4}
Same as Figure \ref{fig3} for the nucleus $^{208}$Pb.}
\end{figure}
Because there is no explicit
radial-dependent effective mass in the full HF calculation of
finite nuclei using the Gogny forces, we compare the DFT-$\hbar^2$
results with the neutron and proton effective mass over the bare
nucleon mass obtained using the Skyrme III force \cite{siii}
(dashed lines). We find that the DFT-$\hbar^2$ results exhibit
similar trends to those of the Skyrme effective masses. The
differences between the two calculations are basically due to the
different values of the nucleon effective mass in nuclear matter
which are $m^*/m$ = 0.70 for the Gogny D1S force and
$m^*/m$ = 0.76 for the Skyrme III interaction.

\section{SUMMARY}

In the present paper we propose a non-local extension of the DFT
and its quasi-local reduction. To this aim we define an energy
functional which depends on the Slater-determinant DM with the
occupation numbers being either 1 or 0. This enables us to avoid the
difficulties of the non-local DFT reported in \cite{Gilb}.
Defining the uncorrelated kinetic-energy densities and spin
densities we construct the quasi-local energy functional and rigorously
derive the single particle equations with the radial-dependent
effective mass and the spin-orbit potential.

In order to define the energy functional of the Slater-determinant
DM one has to introduce an effective Hamiltonian which ensures its
existence. This feature of nucleon systems arises from
the specific properties of the bare NN
force in contrast to the Coulomb force in electron systems.
In our approach the total energy functional consists of
HF part and residual correlation energy. The HF
energy functional can be calculated directly, while the residual
correlation energy is considered phenomenologically.
Using the recently proposed semiclassical ETF approximation
for the DM \cite{SV} we obtain a
quasi-local energy density functional only depending on the local
particle, kinetic-energy and spin densities. The resulting
single-particle equations of motion contain the local mean-field
potential, the uncorrelated effective mass and spin-orbit potential.
Using the finite range density-dependent Gogny force they are
calculated analytically. The use of a different effective 
force such as M3Y will be reported in a forthcoming publication.

We apply our method to the calculations of
the nuclear ground-state properties and compare it with the exact 
HF solutions using the Gogny D1S force. We obtain a very
good agreement in the description of the binding energies and root
mean square radii. The single-particle energies of
the highest occupied
neutron and proton levels in the full HF calculation are well
reproduced by our local approximation. The particle densities are
also in good agreement with the exact HF densities. We analyze the
radial dependent effective mass within our approach. Comparing it
with the result obtained with the Skyrme III interaction, it also exhibits 
a very reasonable behavior.

In conclusion, our approach has the
following advantages: it handles local differential
equations in contrast to the integro-differential equations in the HF
approach, at the same time the
quality of the obtained results is sufficiently high;
our method enables one to construct a quasi-local energy density
functional on the base of the effective
forces with arbitrary radial formfactors;
the method can be straightforwardly generalized to the
non-spherical case.

\section*{Acknowledgments}
The authors are grateful to the Gogny group for providing us the
HF results with the Gogny force and to M. Centelles for a careful
reading of the manuscript. One of us also acknowledges financial
support from DGCYT (Spain) under grant PB98-1247 and from DGR
(Catalonia) under grant 2000SGR-00024.

\newpage
\section*{Appendix 1}

In this appendix we present the explicit expressions for the
functions $F_q$, $G_q$, $F_q^{q'}$, $G_q^{q'}$, $\Delta F_q$ and $\Delta
G_q$
calculated with a Gaussian formfactor. These functions are used
to obtain the second-order contributions to the exchange
nuclear energy density (\ref{eq29}) and the corresponding SPP
(\ref{eq32}). In the following,
excepting Eqs.~(\ref{dfqqq}) and (\ref{dgqqq}), we assume that
$q' \ne q$.
\bea F_q &=& -\frac{a^2}{2\sqrt{\pi}} \{ X_{e1}z_q^3
\exp(-\frac{z_q^2}{2}) Q_1 (\frac{z_q^2}{2})+ X_{e2}z_{q'}^3
\exp(-\frac{z_q^2+z_{q'}^2}{4}) Q_1 (\frac{z_q z_{q'}}{2}) \} \,,
\\
G_q &=& \frac{a^2}{4\sqrt{\pi}}z_q^2\{ X_{e1}z_q^3
\exp(-\frac{z_q^2}{2})[Q_1(\frac{z_q^2}{2})-
z_q^2Q_2(\frac{z_q^2}{2})]\nonumber\\
&+&X_{e2}z_{q'}^3
\exp(-\frac{z_q^2+z_{q'}^2}{4})[Q_1(\frac{z_q z_{q'}}{2})-
z_{q'}^2Q_2(\frac{z_q z_{q'}}{2})]\}\,, \label{eq30}
\eea
where $z_q = a k_{q}$ and
the functions $Q_m(x)$ are defined by
\begin{equation}
Q_0(x) = \frac {\sinh (x)}{x}\,, \qquad
Q_{m+1} = \frac {1}{2x} \frac {d Q_m (x)}{dx}\,.
\end{equation}
The functions $F_q^{q'}$, $G_q^{q'}$, $F_q^{q'q''}$ and $G_q^{q'q''}$
(the two last functions needed to obtain $\Delta F_q$ and $\Delta G_q$,
see below) are defined as:
\be F_q^{q'}
= \frac{1}{k_{q'}}\frac{\partial F_q}{\partial k_{q'}}, \qquad
F_{q}^{q'q''} = \frac{\partial^2 F_q} {\partial k_{q'}
\partial k_{q''}}\,,
\label{dfqqq}
\ee
\be
G_{q}^{q'} = \frac{1}{k_{q'}}\frac{\partial G_q}{\partial k_{q'}},
\qquad G_{q}^{q'q''} = \frac{\partial^2 G_q} {\partial k_{q'}
\partial k_{q''}}\,,
\label{dgqqq}
\ee
and their explicit form is:
\be
F_q^q=-\frac{a^4}{4\sqrt{\pi}}\{2X_{e1}e_qz_q[(3-z_q^2)Q_1(x_q)+
z_q^4Q_2(x_q)]+X_{e2}e_0z_{q'}^3[z_{q'}^2Q_2(x_0)-Q_1(x_0)]\},
\ee
\be
F_q^{q'}=-\frac{a^4}{4\sqrt{\pi}}
X_{e2}e_0z_{q'}[(6-z_{q'}^2)Q_1(x_0)+z_q^2z_{q'}^2Q_2(x_0)],
\ee
\bea
F_q^{qq}&&=-\frac{a^4}{8\sqrt{\pi}}\{4X_{e1}e_qz_q[(2-z_q^2)(3-2z_q^2)
Q_1(x_q)-z_q^4(1+2z_q^2)Q_2(x_q)]\nonumber\\
&&-
X_{e2}e_0z_{q'}^3[(2-z_q^2-z_{q'}^2)Q_1(x_0)
+2z_{q'}^2(4+z_q^2)Q_2(x_0)]\},
\eea
\be F_q^{q'q'}=-\frac{a^4}{8\sqrt{\pi}}
X_{e2}e_0z_{q'}[(24-14z_{q'}^2+z_q^2z_{q'}^2+z_{q'}^4)Q_1(x_0)+
2z_{q}^2z_{q'}^2(2-z_{q'}^2)Q_2(x_0)],
\ee
\be
F_q^{qq'}=\frac{a^4}{8\sqrt{\pi}}
X_{e2}e_0z_qz_{q'}^2[(6-2z_{q'}^2)Q_1(x_0)+z_{q'}^2
(z_q^2+z_{q'}^2)Q_2(x_0)],
\ee
\bea
G_q^q&&=\frac{a^4}{8\sqrt{\pi}}\{2X_{e1}e_qz_q^3[(5-2z_q^2)Q_1(x_q)+
z_q^2(3+2z_q^2)Q_2(x_q)]\nonumber\\
&&+X_{e2}e_0z_{q'}^3[(4-z_q^2-z_{q'}^2)Q_1(x_0)+2z_{q'}^2(3+z_q^2)
Q_2(x_0)]\},
\eea
\be G_q^{q'}=\frac{a^4}{8\sqrt{\pi}}
X_{e2}e_0z_q^2z_{q'}[(6-2z_{q'}^2)Q_1(x_0)+z_{q'}^2
(z_q^2+z_{q'}^2)Q_2(x_0)],
\ee
\bea
G_q^{qq}&&=\frac{a^4}{16\sqrt{\pi}}\{8X_{e1}e_qz_q^3
[(10-7z_q^2+2z_q^4)Q_1(x_q)-
z_q^2(6+z_q^2+2z_q^4)Q_2(x_q)]\nonumber\\
&&+X_{e2}e_0z_{q'}^3[(8-10z_q^2+4z_{q'}^2+3z_q^2z_{q'}^2+z_q^4)
Q_1(x_0)\nonumber\\
&&-z_{q'}^2(48+10z_q^2+z_q^2z_{q'}^2+3z_q^4) Q_2(x_0)]\},
\eea
\bea
G_q^{q'q'}&&=\frac{a^4}{16\sqrt{\pi}}
X_{e2}e_0z_q^2z_{q'}[(24-22z_{q'}^2+3z_{q'}^4+z_q^2z_{q'}^2)Q_1(x_0)
\nonumber\\
&&+z_{q'}^2(4z_q^2+2z_{q'}^2-3z_q^2z_{q'}^2-z_{q'}^4)Q_2(x_0)],
\eea
\bea G_q^{qq'}&&=\frac{a^4}{16\sqrt{\pi}}
X_{e2}e_0z_qz_{q'}^2[(24-6z_q^2-8z_{q'}^2
+3z_q^2z_{q'}^2+z_{q'}^4)Q_1(x_0)\nonumber\\
&&+z_{q'}^2(4z_q^2-6z_{q'}^2-3z_q^2z_{q'}^2-z_q^4)Q_2(x_0)], \eea
where $z_q=a k_q$, $x_q=z_q^2/2$, $x_0=z_p z_n /2$,
$e_q= \exp(-x_q)$, $e_0= \exp(-(x_p+x_n)/2)$.
Finally $\Delta F_q$ and $\Delta G_q$  are given by:
\bea \Delta F_q&&=\frac{\pi^2}{3}\{
\frac{1}{k_q}[(F_q^{qq}-2F_q^q) \frac{(\bnabla
n_q)^2}{n_q}+3F_q^q\Delta n_q]
\nonumber\\
&& +\frac{1}{k_{q'}}[(F_q^{q'q'}-2F_q^{q'}) \frac{(\bnabla
n_{q'})^2}{n_{q'}} +3F_q^{q'}\Delta
n_{q'}]+\frac{6\pi^2}{k^2_q k_{q'}^2}F_q^{qq'}(\bnabla n_q)(\bnabla
n_{q'})\},
\\
\Delta G_q&&=\frac{\pi^2}{3}\{
\frac{1}{k_q}[(G_{q}^{qq}-2G_{q}^{q}) \frac{(\bnabla n_q)^2}{n_q}
+ 3G_q^q\Delta n_q]\nonumber\\
&&+\frac{1}{k_{q'}}[(G_q^{q'q'}-2G_q^{q'}) \frac{(\bnabla
n_{q'})^2}{n_{q'}}+3G_q^{q'} \Delta
n_{q'}]+\frac{6\pi^2}{k^2_q k_{q'}^2}G_q^{qq'}(\bnabla n_q)(\bnabla
n_{q'})\}.
\eea
\newpage

\section*{Appendix 2}

In this Appendix we describe briefly the method to calculate the
centre-of-mass correction to the ground-state energy. As is
well known, the general idea consists of subtracting the
quantity
 \begin{equation}
E^{CM} = <\Psi_{GS}|\frac{{\bf P}^2}{2M}|\Psi_{GS}>
 \label{app1}
 \end{equation}
from $E_{GS}$. Here ${\bf P}$ is the total momentum operator,
$M$ is the total mass of a nucleus. Usually the quantity $E^{CM}$
is represented as a sum of two terms:
 \begin{equation}
E^{CM} = E^{CM}_{1} + E^{CM}_{2}\, ,
 \label{app2}
 \end{equation}
where $E^{CM}_{1}$ is the one-body, $E^{CM}_{2}$ is the two-body
centre-of-mass kinetic energy. The quantity $E^{CM}_{1}$
is defined by formulae:
 \begin{equation}
E^{CM}_{1} = \sum_{q} E^{CM}_{1,q}\,,\qquad
E^{CM}_{1,q} = \frac{1}{2M}
\mbox{Sp} \left( {\bf p}^2 \rho_{q} \right)\,,
 \label{app3}
 \end{equation}
where in accordance with the definition (\ref{defrho})
the following notation is introduced
 \begin{equation}
\rho_{q} = \rho_{q} ({\bf r}, {\bf r}') = \sum_{\sigma}
\rho ({\bf r}, \sigma , q,\; {\bf r}', \sigma , q)\,.
 \end{equation}
Hereinafter the symbol $\mbox{Sp}$ denotes the trace over the
space variables. The subtraction of $E^{CM}_{1}$
leads to the simple renormalization of the
nucleon mass in the single-particle Hamiltonian $h_q$:
$m_{q} \to \bar{m}_{1,q}$,
 \begin{equation}
\bar{m}_{1,q} / m_{q} = M / (M - m_{q})\,.
 \label{app4}
 \end{equation}
The reasonable method for the evaluation of the quantity
$E^{CM}_{2}$ is the Hartree-Fock approximation for the
ground-state wave function $\Psi_{GS}$ in Eq.~(\ref{app1}).
In addition we adopt the following approximation for the
single-particle DM:
 \begin{equation}
\rho ({\bf r}, \sigma , q,\; {\bf r}', \sigma ' , q) =
\frac{1}{2} \delta_{\sigma, \sigma '}
\rho_{q} ({\bf r}, {\bf r}')\,.
 \end{equation}
With these assumptions we have
 \begin{equation}
E^{CM}_{2} = \sum_{q} E^{CM}_{2,q}\,,\qquad
E^{CM}_{2,q} = - \frac{1}{2}
\mbox{Sp} \left( K^{CM}_{2,q} \rho_{q} \right)\,,
 \label{app5}
 \end{equation}
where the single-particle operator $K^{CM}_{2,q}$
is defined as
 \begin{equation}
K^{CM}_{2,q} = \frac{1}{2M} {\bf p}\rho_{q}{\bf p}\,.
 \label{app6}
 \end{equation}
In contrast to the one-body contribution, the subtraction
of $E^{CM}_{2}$ leads to the additional non-locality in the
$h_q$ because in the self-consistent
approach we have to add the non-local operator
$K^{CM}_{2,q}$ to the single-particle kinetic-energy
operator. So the total correction to the $h_q$ is:
 \begin{equation}
\frac{{\bf p}^2}{2 m_{q}} \to
\frac{{\bf p}^2}{2 \bar{m}_{1,q}} +
K^{CM}_{2,q}\,.
 \label{app7}
 \end{equation}
In the local or quasi-local DFT and in similar approaches
the non-locality of $K^{CM}_{2,q}$ in Eq.~(\ref{app7})
leads to unpleasant difficulties. So we use the
simplified method, proposed in Ref.~\cite{Ts},
to take into account the contribution of the operator
$K^{CM}_{2,q}$ in the single-particle equations.

Let us write the density matrix $\rho_q$ in the form:
 \begin{equation}
\rho_q ({\bf r},\, {\bf r}') =
2 \int \frac{d{\bf k}}{(2 \pi)^3}
\bar{n}_q ({\bf R},\, {\bf k})
e^{i {\bf k} {\bf s}}\,,
 \label{app8}
 \end{equation}
where ${\bf R} = ({\bf r} + {\bf r}')/2$,
${\bf s} = {\bf r}' - {\bf r}$,
$\bar{n}_q ({\bf R},\, {\bf k})$
is the momentum distribution function. The approximation
consists of replacing function
$\bar{n}_q ({\bf R},\, {\bf k})$
in Eq.~(\ref{app8}) by some effective constant value
$\bar{n}^{CM}_q$.
Substituting Eq.~(\ref{app8}) with
$\bar{n}_q ({\bf R},\, {\bf k}) = \bar{n}^{CM}_q$
into Eq.~(\ref{app6}) we get
 \begin{equation}
\tilde{K}^{CM}_{2,q} = \bar{n}^{CM}_q \frac{{\bf p}^2}{M}\,.
 \label{app9}
 \end{equation}
The value of $\bar{n}^{CM}_q$ is defined by
the substitution of
$\tilde{K}^{CM}_{2,q}$ into Eq.~(\ref{app5}) instead of
$K^{CM}_{2,q}$. Taking into account Eqs.~(\ref{app3}) and
(\ref{app9}) we obtain:
 \begin{equation}
\bar{n}^{CM}_q = - E^{CM}_{2,q} / E^{CM}_{1,q}\,.
 \label{app10}
 \end{equation}
The quantities $E^{CM}_{1,q}$ and $E^{CM}_{2,q}$
are defined everywhere by
Eqs.~(\ref{app3}), (\ref{app5}) which can be rewritten in the
following forms making use of Eq.~(\ref{rhosl}) for the
Slater-determinant density matrix
 \begin{eqnarray}
E^{CM}_{1,q} & = & \frac{\hbar ^2}{2M}
\sum_{i=1}^N \sum_{\sigma}
\int d{\bf r} |\bnabla \varphi_i
({\bf r},\, \sigma,\, q)|^2\,,
 \label{app11}\\
E^{CM}_{2,q} & = & - \frac{\hbar ^2}{2M}
\sum_{i=1}^N \sum_{i'=1}^N
|\sum_{\sigma} \int d{\bf r}
\varphi_i^* ({\bf r},\, \sigma,\, q)
\bnabla \varphi_{i'} ({\bf r},\, \sigma,\, q)|^2\,.
 \label{app12}
 \end{eqnarray}
It is easy to prove, using Eqs.~(\ref{app11}), (\ref{app12})
and the completeness of the set of functions $\varphi_i$,
that $|E^{CM}_{2,q}| < |E^{CM}_{1,q}|$ and consequently:
 \begin{equation}
0 < \bar{n}^{CM}_q < 1\,.
 \label{app13}
 \end{equation}
 One can also prove that in the limit cases
($\bar{n}^{CM}_q \to 0$ and $\bar{n}^{CM}_q \to 1$)
the action of the approximate operator
$\tilde{K}^{CM}_{2,q}$ upon the
wave functions of the occupied orbitals $\varphi_i$
coincides with the action of the initial operator
$K^{CM}_{2,q}$ as defined by Eq.~(\ref{app6}). So
Eq.~(\ref{app9}) can be considered as the interpolation
formula.

     The total centre-of-mass correction to the $h_q$
in this method is reduced to the renormalization of the
nucleon mass as in the one-body case:
 \begin{equation}
\frac{{\bf p}^2}{2 m_q} \to
\frac{{\bf p}^2}{2 \bar{m}_q}\,,\qquad
\frac{\bar{m}_q}{m_q} = \frac{M}{M
+ (2 \bar{n}^{CM}_q - 1) m_q}\,.
 \label{app14}
 \end{equation}
\newpage

\begin{thebibliography}{99}
\bibitem{Vauth}
D. Vautherin and D. M. Brink,
Phys. Rev. C {\bf 5}, 626 (1972).
\bibitem{Chab}
E. Chabanat, P. Bonche, P. Haensel, J. Meyer, and
R. Schaeffer, Nucl. Phys. {\bf A635}, 231 (1998).
\bibitem{Gog}
J. Decharg\'e and D. Gogny, Phys. Rev. C {\bf 21}, 1568 (1980).
\bibitem{Hof}
F. Hofmann and H. Lenske, Phys. Rev. C {\bf 57}, 2281 (1998).
\bibitem{CB}
X. Campi and A. Bouyssy, Phys. Lett. B {\bf 73}, 263 (1978);
Nukleonica {\bf 24}, 1 (1979).
\bibitem{NV}
J. W. Negele and D. Vautherin, Phys. Rev. C {\bf 5},
 1472 (1972); {\bf 11}, 1031 (1975).
\bibitem{RS}
P. Ring and P. Schuck, The Nuclear Many-Body Problem
(Springer-Verlag, New York Inc., 1980).
\bibitem{SV}
V. B. Soubbotin and X. Vi\~nas, Nucl. Phys.
{\bf A665}, 291 (2000).
\bibitem{KS}
W. Kohn and L. J. Sham, Phys. Rev. {\bf 140},
A1133 (1965).
\bibitem{HK}
P. Hohenberg and W. Kohn, Phys. Rev. {\bf 136},
B864 (1964).
\bibitem{JG}
R. O. Jones and O. Gunnarsson, Rev. Mod. Phys. {\bf 61},
689 (1989).
\bibitem{DFT}
H. Eschrig,
The Fundamentals of Density Functional Theory
(B. G. Teubner, Stuttgart, 1996).
\bibitem{Gilb}
T. L. Gilbert, Phys. Rev. B {\bf 12}, 2111 (1975).
\bibitem{Lieb}
E. H. Lieb, Int. J. Quantum Chem. {\bf 24},
243 (1983).
\bibitem{SV1}
V. B. Soubbotin, X. Vi\~nas, Ch. Roux, P. B. Danilov, and
K. A. Gridnev, J. Phys. G {\bf 21}, 947 (1995).
\bibitem{Ts}
V. I. Tselyaev, Izvestiya Russ. Akad. Nauk, Ser. Fiz.
{\bf 58}, No.5, 61 (1994) [ Bull. Russ. Acad. Sci., Phys. (USA)
{\bf 58}, 762 (1994) ].
\bibitem{DG}
R. M. Dreizler and E. K. U. Gross, Z. Phys. A {\bf 302},
103 (1981).
\bibitem{BGG}
J. F. Berger, M. Girod and D. Gogny, Comput. Phys. Commun.
{\bf 63}, 365 (1991); Nucl. Phys. {\bf A502}, 85c (1989).
\bibitem{Mart}
M. N. Butler, D. W. L. Sprung, and J. Martorell,
Nucl. Phys. {\bf A422}, 157 (1984).
\bibitem{BBDG}
J. P. Blaizot, J. F. Berger, J. Decharg\'e, and M. Girod,
Nucl. Phys. {\bf A591}, 435 (1995).
\bibitem{siii}
M. Beiner, H. Flocard, Nguyen Van Giai, and P. Quentin, Nucl.
Phys. {\bf A238}, 29 (1975).             
\end{thebibliography}
\end{document}